\documentclass[prl,twocolumn]{revtex4-2}
\newcommand{\beq}{\begin{equation}}
\newcommand{\eeq}{\end{equation}}
\usepackage{graphicx}
\usepackage{color}
\begin{document}
\title{
Circular Dichroism in Resonant Photoelectron Diffraction
as a Direct Probe of Sublattice Magnetization in Altermagnets
}
\author{Peter Kr\"{u}ger}
\affiliation{Graduate School of Engineering
and Molecular Chirality Research Center,
Chiba University, Chiba 263-8522 Japan}

\date{25 September 2025}

\begin{abstract}
Altermagnets are a new class of magnetic materials that are promising
for spintronics technology. Here it is shown that in altermagnets,
the circular dichroism (CD) in resonant photoelectron diffraction (RPED) 
contains a time-reversal odd signal, which provides a direct probe of the
sublattice magnetization.
RPED calculations are performed for MnTe at the Mn L$_{2,3}$-edge resonance,
using a combination of atomic multiplet and multiple scattering theory.
A large magnetic CD is found for light helicity parallel to the N\'eel vector.
This signal has the same angular distribution as the difference between
the structural RPED of the two magnetic sublattices and its amplitude
is approximately proportional to the x-ray magnetic CD in absorption of a
single sublattice, thus providing a direct probe of the local magnetic
moments.
\end{abstract}

\maketitle
Altermagnets (AMs) are a new class of magnetic
materials~\cite{hayami19,libor20,libor22} with great potential 
for spintronics applications, among other reasons
because AMs combine the possibility of spin-dependent
currents~\cite{chou24} with fast magnetic switching~\cite{jungwirth16,han24}. 
An AM has a compensated spin structure with
two equivalent magnetic sublattices of opposite spin.
In this respect, AMs are a special class of antiferromagnets.
In a conventional antiferromagnet, the two sublattices are connected by an
inversion or a translation symmetry operation. 
All band states are spin-degenerate and so the carriers cannot be
spin-polarized.
In AMs, the two sublattices are connected by a
different symmetry operation, such as a rotation or a screw
axis~\cite{libor22}. As a consequence, spin-splitting of certain
band states occurs~\cite{krempasky24} and spin-polarized transport
may become possible.
While the knowledge of the space group of a candidate material is
sufficient for deciding whether an altermagnetic phase may exist in principle,
the experimental proof of the altermagnetic state is difficult.
Even for ruthenium dioxide, one of the first and most studied AM candidate
materials, the existence of the altermagnetic phase is still
controversial~\cite{libor20,kessler24}.
There is a need for experiments that can reliably determine
the microscopic magnetic structure of AM candidate materials.
Neutron scattering is bulk sensitive and cannot be used for
heterostructures and thin films which are particularly important
for device applications.
X-ray magnetic circular dichroism (XMCD), on the other hand,
can be used for a wide range of systems, including molecules,
nanomaterials and surfaces. It is element-selective and the local
spin- and orbital moment of the magnetic atoms can be directly inferred
from experimental spectra via the XMCD sum rules\cite{thole92,carra93}.
The XMCD produced by a single magnetic atom is largest when the helicity
vector ${\bf q}$ of the light is parallel to the magnetic moment.
In an altermagnet with N\'eel vector ${\bf L}$, the magnetic atoms on sites A
(B) sites have a local moment per atomic volume of
${\bf L}$ ($-{\bf L}$). As there are as many A sites as B sites,
the total parallel XMCD exactly vanishes.
Perpendicular XMCD, on the other hand, may be non-zero in AMs.
Indeed, in MnTe with ${\bf L}\sim [1{\bar 1}00]$ and light
incoming along the crystal $[0001]$ axis,
a small net XMCD signal was observed by Hariki et al.~\cite{hariki24}.
This perpendicular XMCD effect is, however, by over one order of magnitude
weaker than the parallel XMCD of a single sublattice and it is unrelated
to the magnetic dipole moments. Moreover, it was shown that if ${\bf L}$
were oriented along [11${\bar 2}$0] in MnTe,
then the perpendicular XMCD would vanish by symmetry~\cite{hariki24}.
This suggests that perpendicular XMCD is not a generally applicable technique
for identifying altermagnetism in candidate materials.
A weak parallel XMCD signal has been predicted theoretically for the
altermagnet candidates RuO$_2$~\cite{sasabe23,hariki24b}
and NiF$_2$~\cite{hariki25}. This parallel XMCD
is either due to weak ferromagnetism~\cite{hariki25}, or to the anisotropy
of the magnetic dipole operator~\cite{sasabe23,hariki24b}.
In both cases, it is not directly related
to the macroscopic staggered magnetisation, which is the main order parameter
of compensated magnets, and which can be measured with the present method.

In this letter, we show theoretically, that resonant photoelectron
diffraction (RPED) effectively
combines the magnetic information of XMCD with local structural
information from photoelectron diffraction (PED) which, in the case of
AMs, makes it possible to directly probe the local magnetic moments.
X-ray PED is a technique for element- and site-resolved structural
characterization at crystalline surfaces~\cite{fadley84}.
In RPED from valence states excited at a core-valence resonance,
the structural information from PED is augmented with local electronic
information from both occupied and unoccupied valence states.
Thereby, valence band photoemission data can be decomposed into contributions
from different atomic sites near the surface~\cite{kruger08}.
Morscher et al.~\cite{morscher11} showed that for a ferromagnetic surface,
the magnetization direction can be inferred from the circular
dichroism (CD) signal in RPED, which is a direct consequence of
the XMCD in the x-ray absorption step of the resonant photoemission
(RPES) process~\cite{sagehashi23}.
However, CD also occurs in RPED from non-magnetic systems~\cite{matsui15}.
Such CD in angular distribution is a photoemission final state effect
due to scattered wave interference,
first observed in core-level PED by Daimon et al.~\cite{daimon93}.
Let us note that spin-resolved RPES was suggested as a probe
for local magnetic moments in antiferromagnetic and magnetically
disordered systems~\cite{sinkovic97}
but the interpretation is controversial~\cite{vdlaan98,dapieve13}.

Here we report valence-band RPES and RPED calculations 
from a MnTe(0001) surface with photon energies around the
Mn-L$_{2,3}$ absorption edges.
We find that the RPED patterns display a large magnetic circular dichroism
(MCD) in parallel geometry.
The photon energy dependence of the MCD is approximately proportional to the XMCD of a single
magnetic sublattice and thus provides a direct measure of the local magnetic
moments.
The effects found here for MnTe are general for AM materials,
establishing CD-RPED as a novel, powerful technique for the
magnetic characterization of AM surfaces and thin films.

\begin{figure}
\includegraphics[width=\columnwidth]{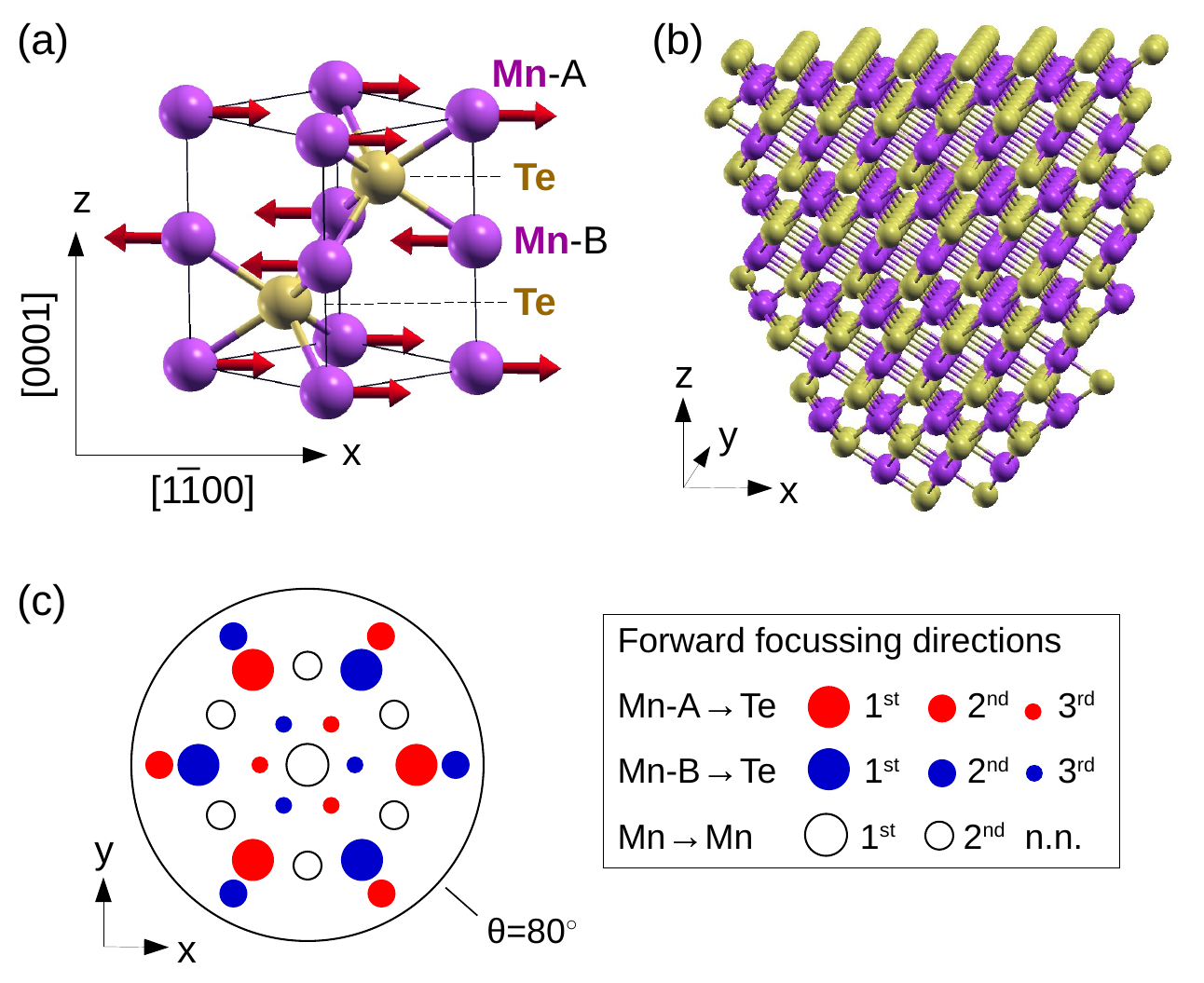}
\caption{(a) Ball-and-stick model of MnTe
  in altermagnetic ground state with N\'eel vector ${\bf L}\sim x$.
(b) Cluster model of Te/Mn-A terminated (0001) surface.
  (c) Mayor forward focussing directions in stereographic projection.
  Here and in all other figures the maximum polar angle is $\theta=80^\circ$. }
  \label{fig1}
\end{figure}
The crystal structure of $\alpha$-MnTe is shown in Fig. 1(a).
The unit cell contains two Mn sites denoted A and B, which are equivalent
by a $6_3$ screw axis. In the altermagnetic ground state, the N\'eel vector
${\bf L}$ is along $[1{\bar 1}00]$ which is taken as the $x$-axis.
In the RPED calculations we consider a Te-terminated MnTe(0001)
surface. There are two possibilities for the subsurface layer,
Mn-A and Mn-B. Since a real surface unavoidably contains steps,
we average all results over these two cases.
We note that details of the MnTe(0001) surface structure are not fully
known and may depend on sample preparation.
However, the conclusions of this work are independent of the surface
structure. Indeed, the main finding, namely the existence of a large
MCD in the RPED pattern, is a bulk phenomenon as is evident from
symmetry considerations and as we have checked
by RPED simulations for bulk MnTe, see S.M.~\cite{supp}.


The RPED calculations are done with our recent method~\cite{sagehashi23},
which combines crystal field multiplet theory~\cite{kruger20}
for the resonant photoemission process and real space multiple
scattering theory for the propagation of the photoelectron
wave~\cite{fadley84,edac}.
For the latter, finite clusters with 479 atoms are used, see Fig. 1(b).
For given photon energy and light helicity, a RPED calculation is performed
for each of the 210 final state multiplets of the Mn($3d^4$) configuration,
with the photoelectron source wave
obtained from the multiplet calculation, and the RPED intensities are summed.
Further theoretical and numerical details are given in the S.M.~\cite{supp}.

\begin{figure}
\includegraphics[width=\columnwidth]{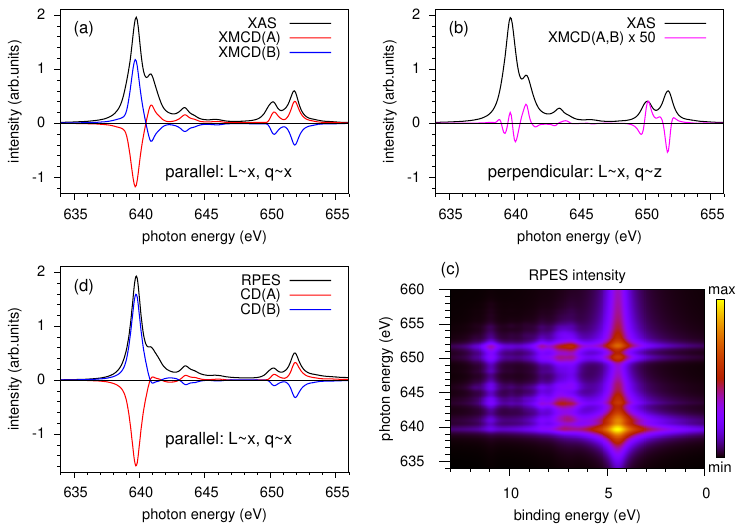}
\caption{Multiplet calculation for MnTe.
  (a) Mn L$_{2,3}$-edge XAS and XMCD spectra in parallel geometry,
  i.e. light helicity ${\bf q}$ $\|$ ${\bf L}\sim x = [1{\bar 1}00]$.
  XMCD(S) is the atomic XMCD for site S=A,B.
  (b) XAS and XMCD in perpendicular geometry with ${\bf q}\sim z=[0001]$.
  The XMCD signal is multiplied by a factor 50.
  (c) Resonant photoemission spectra (RPES) in parallel geometry
  as a 3D color plot.
  The color scale is linear in $\log(1+100\times I)$
  where $I$ is the RPES intensity.
  (d) Total, binding energy integrated RPES intensity and its CD.
} \label{fig2}
\end{figure}
Figures 2(a) and 2(b) show the x-ray absorption spectroscopy (XAS)
and XMCD spectra
obtained with the crystal field multiplet model.
XAS (XMCD) is the sum (difference) spectra of light with positive
(+) and negative helicity ($-$).
In Fig. 2(a), for parallel geometry with ${\bf q}$$\sim$$x$,
XMCD(A) and XMCD(B) are the local signals from single sites Mn-A and Mn-B,
respectively.
The XMCD is large and has a lineshape typically for a
high-spin Mn$^{2+}$ ion~\cite{hariki24}.
XMCD(B) is exactly the negative
of XMCD(A), so the total XMCD vanishes.
In perpendicular geometry with ${\bf q}$$\sim$$z$ [Fig. 2(b)],
there is a very small XMCD signal (note the scaling factor $\times 50$)
with fast oscillations.
This signal is the same for sublattices A and B, leading
to a non-zero total XMCD. The spectra in Figs.~2(a) and 2(b)
agree very well with Ref.~\cite{hariki24},
validating the present crystal field multiplet model.

The RPES spectra for ${\bf L}$, ${\bf q}\sim x$,
are plotted in Fig.~2(c) as a function of photon and binding
energy with a logarithmic color scale for the intensity.
There is a strong resonance at the absorption L$_3$-edge and a
weaker one at the L$_2$-edge.
The photoemission line shape depends rather weakly on the photon energy,
although high binding energy features (e.g. the peak at 11 eV)
become stronger for higher photon energy.
This reflects the fact that for intermediate states
above the L$_3$ absorption threshold, many-body ``shake-up''
excitations become possible which shift spectral weight from low to high
binding energy.

The total RPES intensity, integrated over binding energy,
is shown in Fig.~2(d).
CD(A) and CD(B) are the RPES-CD spectra for sites A and B, respectively.
The CD(A) lineshape is very similar to XMCD(A) in Fig.~2(a) and
the normalized RPES-CD
is approximately equal to the normalized XMCD, i.e.
\begin{equation}\label{icd}
  I_{\rm CD(A)}(\omega)/I_{\rm RPES}(\omega)
  \approx I_{\rm XMCD(A)}(\omega)/I_{\rm XAS}(\omega)
\end{equation}
where $\omega$ is the photon energy and $I$ is the intensity.
The small differences between the lineshapes in Figs.~2(a) and 2(d)
are attributed to the dependence of the Auger decay probability
on the intermediate multiplet state ($2p^53d^6$).
Importantly, however, the CD in RPES displays the same sign changes as XMCD,
when going from site A to B or from the L$_3$-edge at $\sim$640~eV
to the L$_2$-edge at $\sim$652~eV.
This means that the local CD-RPES signal for site A or B,
contains essentially the same magnetic information as the local XMCD.
In XMCD and angle-integrated RPES experiments, such
site-resolved information is not accessible,
since the signals of A and B are of opposite sign and cancel out.
However, site-resolution can be achieved in a RPED experiment,
because the PED patterns for Mn-A emission and Mn-B emission are different.
For MnTe, this will be shown numerically below and may be anticipated
from the forward focussing directions in Fig.~1(c).
Importantly, this argument is valid for all altermagnets,
because the rotation part (in MnTe: rotation by $\pi$ around $z$)
of the space group operation between the two magnetic sublattices
(in MnTe: screw axis $6_3$)
cannot be a symmetry operation of the point group of the magnetic
sites (in MnTe: $S_6$)~\cite{libor22}.
Put more simply, the two magnetic sites A and B have differently oriented
crystal environments and so their diffraction patterns are distinct.

\begin{figure}
\includegraphics[width=0.95\columnwidth]{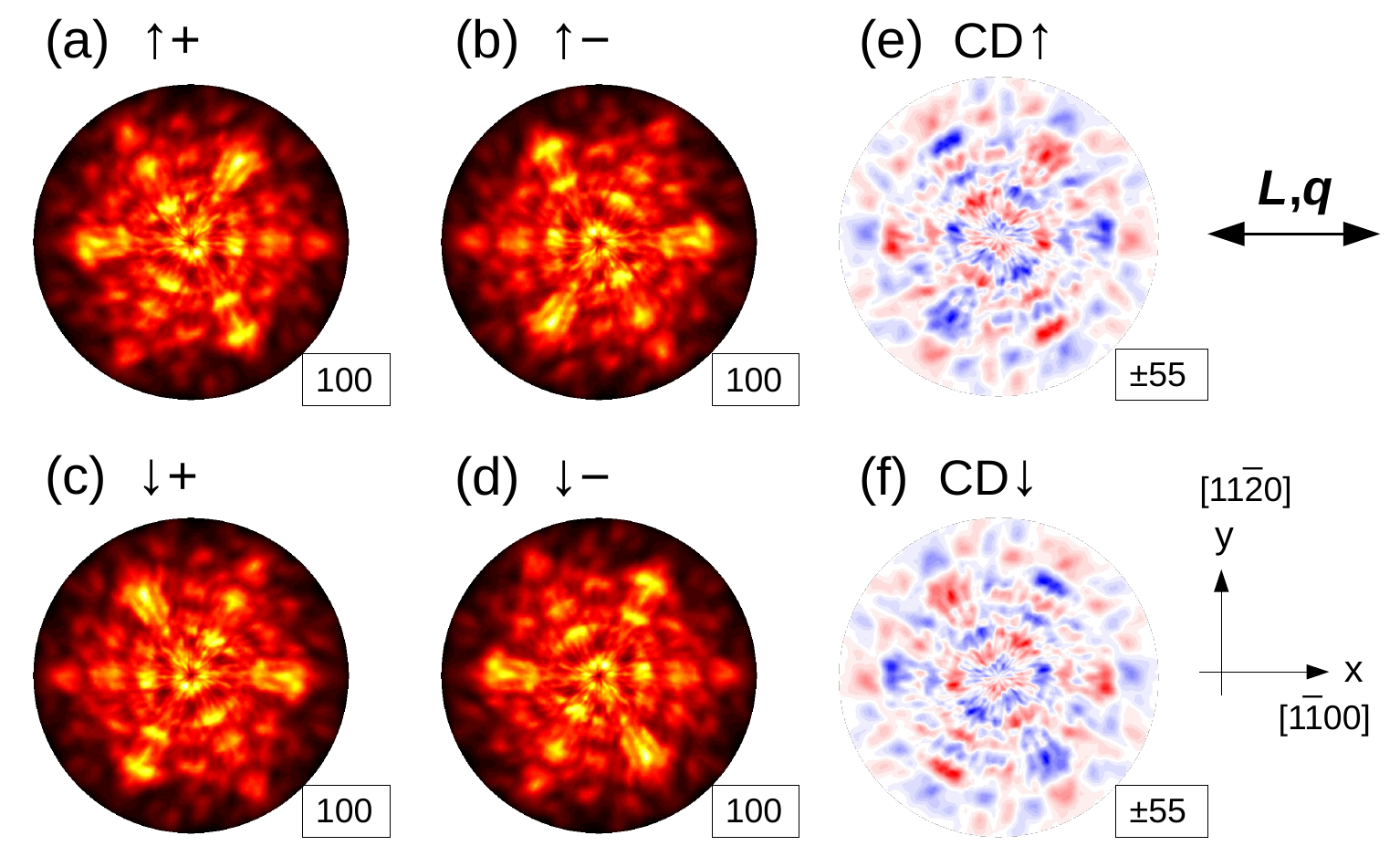}
\caption{Calculated resonant photoelectron diffraction patterns
of MnTe(0001) at the Mn-L$_3$ resonance ($\hbar\omega=639.7$~eV).
Light is incoming along $[1{\bar 1}00]$ with positive ($+$) or negative
($-$) helicity, as indicated.
Two altermagnetic domains are considered with N\'eel vector ${\bf L}$
parallel ($\uparrow$) or antiparallel ($\downarrow$) to $[1{\bar 1}00]$.
All plots are a stereographic projections for polar angles up to $80^\circ$.
In (a-d) the maximum intensity is yellow and minimum is black.
In the CD patterns (e,f) positive (negative) values are red (blue).
The small boxes indicate the maximum intensity relative to that of (a) which
is put to 100.
} \label{fig3}
\end{figure}
For all RPED results shown in this paper, the photoemission
intensity has been integrated over the whole valence band.
The binding energy dependence of RPED contains
information about the wave function character of the valence
states~\cite{kruger08,sagehashi23} but this will not be discussed here.
RPED patterns for a photon energy of 639.7~eV, 
at the maximum L$_3$ resonance, are shown in Fig.~3(a)-3(d),
for parallel geometry
with ${\bf q}$ and ${\bf L}$ parallel to $[1{\bar 1}00]$
and both signs of ${\bf L}$
($\uparrow$, $\downarrow$)
and ${\bf q}$ ($\pm$).
All patterns have approximate 3-fold symmetry.
Considering the hexagonal space group of MnTe (P6$_3$/mmc)
one might expect a diffraction pattern with 6-fold symmetry.
When the four patterns in Figs.~3(a)-3(d) are added up, i.e. light helicity and
magnetization directions are averaged, then the resulting pattern has
indeed 6-fold symmetry, see S.M.~\cite{supp}.

The finding that the individual patterns in Figs.~3(a)-3(d) are 3-fold,
can be understood as follows.
From Fig.~2(d) we know that at the L$_3$-resonance,
there is a large negative (positive) CD in RPES
for emission from Mn-A (Mn-B).
As a consequence, for ($\uparrow$$+$), i.e. ${\bf L}\sim +x$ and positive helicity,
the RPES intensity is much larger for Mn-B than Mn-A, and thus
 ($\uparrow$$+$) looks essentially like a PED pattern from site Mn-B.
Indeed, the strongest peaks in ($\uparrow$$+$) are found at the
first nearest neighbor (nn) forward focussing directions Mn-B$\rightarrow$Te,
as seen by comparison with Fig.~1(c).
When reversing the light helicity, we obtain pattern ($\uparrow$$-$) in
Fig.~3(b), which is the $\pi$-rotated image of ($\uparrow$$+$).
In ($\uparrow$$-$), emission from Mn-A sites dominates.
When reversing both light helicity and N\'eel vector we obtain the pattern
($\downarrow$$-$) which is a mirror image of ($\uparrow$$+$) w.r.t.
the $x$-axis in the plot, i.e. w.r.t. the (11${\bar 2}$0) plane.
The patterns ($\uparrow$$+$) and ($\downarrow$$-$) are quite similar.
This is expected because the relative orientation of light helicity and N\'eel
vector is the same and so the atomic XMCD effect on each site is unchanged.
We conclude that in RPED with circular polarized light,
the relative orientation of helicity and N\'eel vector determines
the Mn sublattice from which the majority of photoelectrons are emitted.
When taking the difference of ($\uparrow$$+$) and ($\uparrow$$-$)
the pattern CD$\uparrow$ in Fig.~3(e) is obtained.
The CD is large, with a maximum contrast of $\pm 55$\%.
Upon reversal of ${\bf L}$ 
[CD$\downarrow$, Fig.~\ref{fig3}(f)] most features change sign,
which shows that the CD is mostly a magnetic effect.

However, since in angle-resolved photoemission,
CD is common even at non-magnetic surfaces,
it is necessary to disentangle the structural and the magnetic CD.
The structural CD
can be obtained by summing the patterns CD$\uparrow$
and CD$\downarrow$. This corresponds to a demagnetized sample with
randomly oriented domains.
Here, the structural CD is clearly non-zero~\cite{supp}.
When the patterns CD$\uparrow$ and
CD$\downarrow$ are subtracted, any structural CD cancels,
and we obtain the purely magnetic CD, given by
\begin{equation}\label{mcdef}
{\rm MCD}
= I(\uparrow+)-I(\uparrow-)-I(\downarrow+)+I(\downarrow-)
\end{equation}
where $I$ denotes the RPED intensity.
Upon time-reversal, ${\bf L}\rightarrow -{\bf L}$,
or $(\uparrow)\rightarrow(\downarrow)$ in our notation,
MCD changes sign, i.e. MCD is time-reversal (${\cal T}$) odd.
This implies that in a ${\cal T}$-even system, including paramagnets and
conventional antiferromagnets, MCD vanishes exactly.
A non-zero MCD signal is a direct signature of ${\cal T}$-symmetry breaking
and thus provides an easy way to distinguish between altermagnets and
conventional antiferromagnets.
To check this point explicitly in an example, we calculated CD-RPED for the
conventional antiferromagnet MnO, and found MCD=0, as expected~\cite{supp}.

Note that by changing the signs of both helicity and N\'eel vector
four independent, ``fundamental'' signals can be formed by linear
combination.
Apart from the structural CD and the MCD (Eq.~\ref{mcdef}),
we define a ``structural RPED'' (sum of all 4 patterns)
and a ``magnetic RPED''~\cite{supp}.

\begin{figure}
\includegraphics[width=\columnwidth]{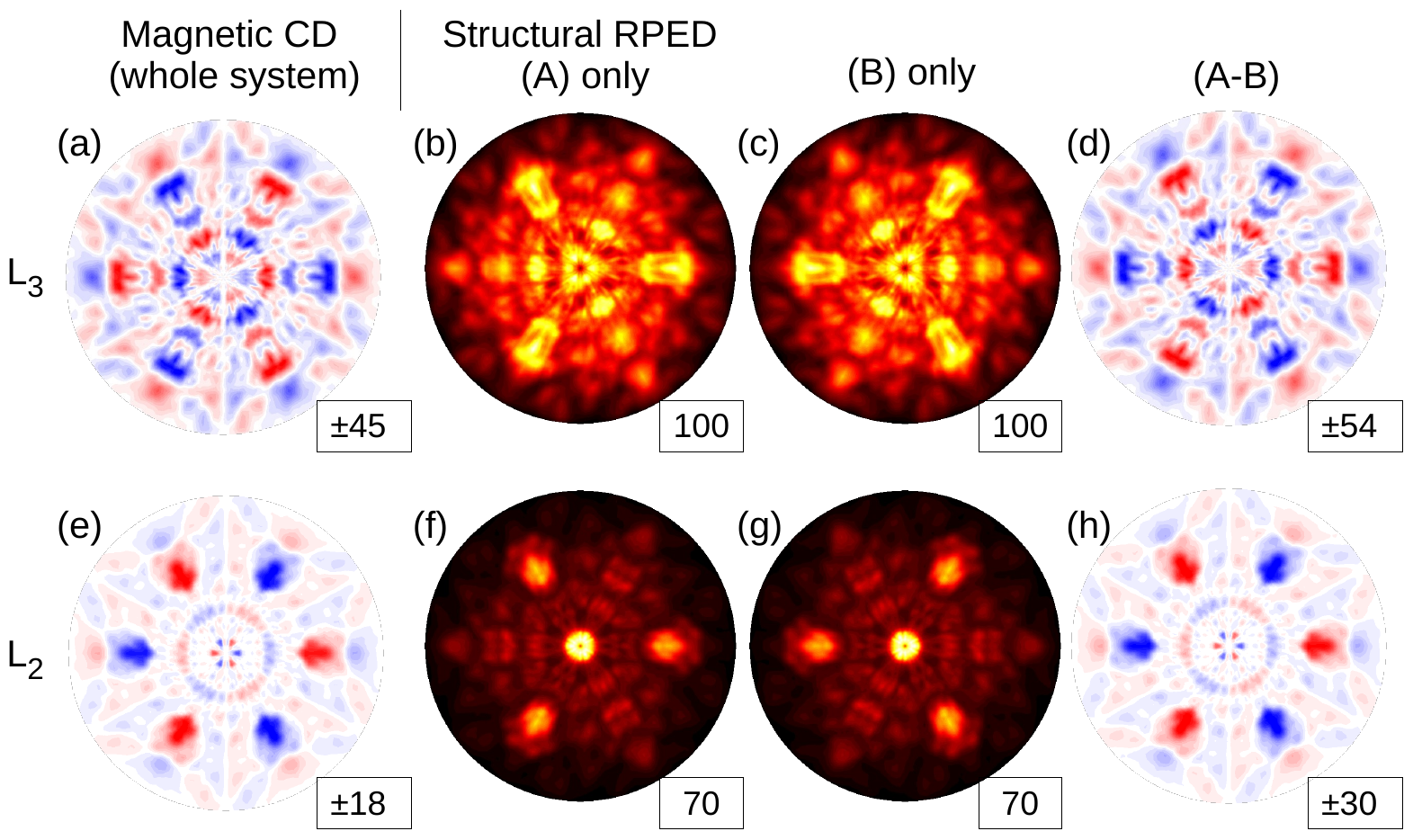}
\caption{Calculated RPED patterns of MnTe for circular polarized light
  with helicity ${\bf q}$ $\|$ ${\bf L}\sim [1{\bar 1}00]$ at the
  L$_3$-resonance ($\hbar\omega$$=$$639.7$eV, a-d) or the L$_2$-resonance
  ($\hbar\omega$$=$$651.8$eV, e-h).
  (a) Magnetic CD pattern, i.e.
  the difference between CD$\uparrow$ and CD$\downarrow$
  in Figs.~3(e) and 3(f).
  (b) Structural RPED for sublattice Mn-A, defined as the sum of patterns
  ($\uparrow$$+$), ($\uparrow$$-$), ($\downarrow$$+$) and ($\downarrow$$-$),
  but restricted to emission from sites Mn-A.
  (c) Same as (b) but for sublattice Mn-B. (d) Difference between 
 (b) and (c). (e-h) Corresponding patterns for the L$_2$-resonance.
  The color scale is always from minimum to maximum. The maximum intensity
  is indicated in the small boxes, relative to that of (b) which is put to 100.
} \label{fig4}
\end{figure}

The MCD pattern shown in Fig.~\ref{fig4}(a) is clearly non-zero,
proving that ${\cal T}$-symmetry is broken in MnTe.
The MCD pattern has the full $S_6$ point symmetry
of the Mn sites. The symmetry of the pattern is not
lowered by ${\bf L}$ or ${\bf q}$ although these vectors
are oriented along the non-symmetry axis~$x$.
We have also computed the RPED-CD by assuming ${\bf L}$ and ${\bf q}$
be oriented along the $z$-axis~\cite{supp}.
Remarkably, the MCD pattern is exactly the same as that in
Fig.~\ref{fig4}(a).
This shows that the MCD in parallel geometry is independent
of the absolute orientation of the N\'eel vector ${\bf L}$
but only depends on the {\em relative}\/ orientation between ${\bf L}$ and
helicity ${\bf q}$ in form of the scalar product ${\bf L}.{\bf q}$.
This corroborates the conclusion that the MCD in RPED is a direct
consequence of the XMCD on each ferromagnetically ordered sublattice.
Indeed, XMCD in a ferromagnet also depends only on the
scalar product between the magnetization~${\bf M}$ and ${\bf q}$
rather than on the absolute orientation of~${\bf M}$. 

Figure~\ref{fig4}(e) shows the MCD in RPED, obtained at the maximum
L$_2$ resonance (651.8~eV).
The pattern has the same symmetry as that at the L$_3$-resonance
[Fig.~4(a)] but the intensity distribution is quite different.
This is because angular distribution of the RPES
source waves strongly depends on the total angular momentum~$j$
of the 2$p$-core hole in the intermediate state.
For both the L$_3$ and the L$_2$ MCD pattern,
the main peaks appear at forward focusing directions [see Fig.~1(c)],
but the sign of the CD is reversed between L$_3$ and L$_2$.
So the MCD signal in RPED changes sign in the same way 
as the sublattice XMCD [Fig.~2(a)] when going from the $L_3$-edge
to the $L_2$-edge.

To better understand the angular distribution of the MCD pattern,
we have computed ``structural RPED'' patterns, defined as
($\uparrow$$+$)+($\uparrow$$-$)+($\downarrow$$+$)+($\downarrow$$-$),
but restricted to emission from one sublattice (Mn-A or Mn-B).
These patterns [Figs.~4(b), 4(c), 4(f), and 4(g)] reflect the crystal structure
around a site Mn-A or Mn-B, and they are dominated by
forward focussing peaks as expected from Fig.~1(c).
Since the N\'eel vector is averaged over
opposite directions, these patterns contain no magnetic information.
The difference between the patterns in Figs.~4(b) and 4(c)
is shown as (A$-$B) in Fig.~4(d).
Quite surprisingly, the purely structural RPED pattern (A$-$B) is identical,
up to a constant scaling factor of $R\equiv I_{\rm MCD}/I_{\rm A-B}=-0.84$,
with the MCD pattern in Fig.~4(a).
This may be explained by the fact that A$-$B (MCD)
can be written as the sum (difference) of two purely structural patterns,
which have the same angular distribution, see S.M. for details~\cite{supp}.
Also for the L$_2$-resonance [Fig.~4(f)-4(h)],
the MCD pattern is exactly the same as (A$-$B),
except for a constant factor $R=+0.61$.

These $R$-values ($-0.84,0.61$) are very close to
the intensity ratio between the angle-integrated CD(A) and RPES
in Fig.~2(d) at corresponding photon energies (639.7~eV and 651.8~eV),
so we have
\begin{equation}\label{mcd}
  \frac{I_{\rm MCD}(\omega,\Omega)}{I_{\rm A-B}(\omega,\Omega)}
 = R(\omega)
  \approx \frac{I_{\rm CD(A)}(\omega)}{I_{\rm RPES}(\omega)}
  \approx \frac{I_{\rm XMCD(A)}(\omega)}{I_{\rm XAS}(\omega)}
\end{equation}
where the function $R(\omega)$ is independent of the
solid angle $\Omega$ and Eq.~(\ref{icd}) has been used.
Fig.~4 and Eq.~(\ref{mcd}) show that the MCD signal,
if normalized by the purely structural signal (A-B),
approximately measures the XMCD of one sublattice.
The quality of this approximation might depend on the material
and deserves further investigation.
Eq.~(\ref{mcd})
suggests that MCD in RPED may be used in conjunction with the XMCD
sum rules~\cite{thole92,carra93} for obtaining
the sublattice magnetization in AMs directly from experiment.
In practice, for obtaining the ratio $R(\omega)$ in Eq.~(\ref{mcd})
one can simulate the purely structural RPED patterns of sublattices (A) and (B) as in Fig.~4(b), 4(c), 4(f), and 4(g).
Importantly, such a simulation is independent of the magnetic state and only requires structural information.
Moreover, patterns (A) and (B) are related by a known symmetry operation (a rotation by $\pi$ around $z$ in the case of MnTe).
The intensity of the sum (A+B) must be normalized to the experimentally measured signal
(A+B)=($\uparrow$$+$)+($\uparrow$$-$)+($\downarrow$$+$)+($\downarrow$$-$).
Then, the ratio $R$ is easily obtained as the best-fit proportionality factor between the measured MCD and the simulated (A$-$B) pattern.

We have defined the MCD in RPED as the difference between two CD patterns
for opposite orientations of the N\'eel vector
(CD$\uparrow$ and CD$\downarrow$)
i.e. the CD patterns of two altermagnetic domains. 
Experimentally, it might be difficult to prepare samples for
different magnetic domains.
In this case, instead of CD$\downarrow$, one
can measure the CD of the demagnetized sample, which is the average
of CD$\uparrow$ and CD$\downarrow$. The pattern CD$\downarrow$
can then be recovered by subtraction.
In a multidomain sample, the MCD signal is proportional to
the average of $\cos\theta$, where $\theta$ is the angle between the local
N\'eel vector ${\bf L}$ and the light helicity~${\bf q}$. This follows
directly from the corresponding angular dependence of the XMCD of
one sublattice.
Therefore, MCD in RPED can in principle be used to map the distribution
of the N\'eel vector~\cite{amin2024}.

In summary, we have presented a theory of RPED from altermagnets
and applied it to MnTe at the Mn L$_{2,3}$-edge resonance.
We find a large magnetic CD signal in
the RPED pattern for parallel geometry, where the XMCD in absorption vanishes.
We have shown that the occurrence of magnetic CD in RPED is a direct
consequence of the XMCD on each magnetic sublattice
and the fact that in an AM, the two sublattices necessarily have distinct
PED patterns. 
Upon normalizing the MCD signal with the structural difference
pattern of the two sublattices, it should be possible to extract the
approximate XMCD spectrum of one AM sublattice and thus obtain
the atomic magnetic moments directly from the experimental data.
Our main findings are valid for any AM and lay the ground for
making CD-RPED a powerful technique for the magnetic characterization of AM
candidate materials.

{\it Acknowledgements}---I thank Alberto Marmodoro
and the late Ondrej {\v S}ipr for bringing my attention
to circular dichroism in altermagnets.

{\it Data availability}---The data that support the findings of
this Letter are not publicly available upon publication
because it is not technically feasible and/or the cost of
preparing, depositing, and hosting the data would be
prohibitive within the terms of this research project. The
data are available from the authors upon reasonable request.


\bibliography{mnte}{}

\end{document}


\title{
Supplemental Material for\\
Circular dichroism in resonant photoelectron diffraction
as a direct probe of sublattice magnetization in altermagnets
}
\author{Peter Kr\"{u}ger}
\affiliation{Graduate School of Engineering
and Molecular Chirality Research Center,
Chiba University, Chiba 263-8522 Japan}


\begin{abstract}
  {
The following supplementary information is given.
  (i) Computational details.
}
  (ii) Fundamental RPED patterns are defined and shown for L$_2$- and
  L$_3$-resonance with N\'eel vector along $x$ or $z$. 
  (iii) An explanation is given why the magnetic CD pattern is, up
  to a constant factor, identical with the structural pattern (A$-$B).
  (iv) RPED calculations for bulk MnTe are shown, which
  demonstrates that the magnetic CD in RPED is not a surface effect.
{
  (v) Comparison with a conventional antiferromagnet (MnO).
}
\end{abstract}

\maketitle
{
\subsection*{(i) Computational details}
The numerical calculations have been done with our recent method
for RPED~\cite{sagehashi23}
which combines crystal field multiplet theory~\cite{kruger20}
for the resonant photoemission process and multiple
scattering theory for the propagation of the photoelectron
wave~\cite{fadley84,edac}.
The resonant photoemission amplitude of the emitter atoms are computed
in a ligand field multiplet model using second order perturbation theory
(Kramers-Heisenberg formula)~\cite{fadley84,edac}.
The Mn-2p and 3d level energies are adapted from experiment.
The atomic Coulomb and spin-orbit integrals
are taken from Tanaka and Jo~\cite{tanaka94} for a Mn$^{2+}$ ion and
the crystal field is taken from A. Hariki et al.~\cite{hariki24}.
In order to orient the local magnetic moment to the
experimentally observed $[1{\bar 1}00]$ direction,
a small exchange field of 1~meV is applied.
The $2p$ core-hole lifetime is set to $\Gamma=0.4$~eV~\cite{hariki24}.
The same multiplet model is used to compute the X-ray
absorption and XMCD spectra (Fig.~2a,b in the main text)
and the results agree very well with A. Hariki et al.~\cite{hariki24}.

For computing the RPED pattern, i.e.
the angular distribution of the resonant photoemission intensity,
a photoelectron diffraction (PED) calculation is performed with
the EDAC code~\cite{edac} for each multiplet final state
and the RPED intensities are summed. For a Mn$^{2+}$ ion, there are
210 final states in the ($3d^4$) configuration.
The source waves of these PED calculations are constructed from
the resonant photoemission amplitudes obtained
in the multiplet calculation, see Ref.~\cite{sagehashi23} for
details.
In the PED calculation the inelastic mean free path is set to 1.37 nm,
according to the universal curve~\cite{imfp}.
The multiple scattering series is truncated at 5th order, which is found
sufficient for convergence of the RPED patterns.
The MnTe surface is modeled with finite clusters of 479 atoms
of semi-ellipsoid shape, see Fig.~1b in the main text.
The clusters have a diameter and depth of 2.7 nm,
which is twice the inelastic mean free path.
Two surface terminations, Te/Mn-A and Te/Mn-B, are considered and
the results are averaged.
}

\subsection*{(ii) Fundamental RPED patterns}
In RPED from an AM with N\'eel vector ${\bf L}$ and
circular polarized light of helicity~${\bf q}$,
there are four possible parallel alignments between
${\bf L}$ ($\uparrow$,$\downarrow$) and ${\bf q}$ ($\pm$), namely
($\uparrow$$+$), ($\uparrow$$-$), ($\downarrow$$+$) and ($\downarrow$$-$).
We define the four ``fundamental'' linear combinations as
\begin{eqnarray}
  {\rm SPD} &=& (\uparrow +)+(\uparrow -)+(\downarrow +)+(\downarrow -) \nonumber \\
  {\rm SCD} &=& (\uparrow +)-(\uparrow -)+(\downarrow +)-(\downarrow -) \nonumber \\
  {\rm MPD} &=& (\uparrow +)+(\uparrow -)-(\downarrow +)-(\downarrow -) \nonumber \\
  {\rm MCD} &=& (\uparrow +)-(\uparrow -)-(\downarrow +)+(\downarrow -) \nonumber
\end{eqnarray}  
where SPD, SCD, MPD and MCD stand for structural RPED,
structural CD, magnetic RPED and magnetic CD, respectively.
These patterns are shown in Fig.~S1 for the Mn L$_3$-edge or L$_2$-edge
with ${\bf L}$, ${\bf q}$ along $x=[1{\bar 1}00]$ or $z=[0001]$.
\begin{figure}
\includegraphics[width=\columnwidth]{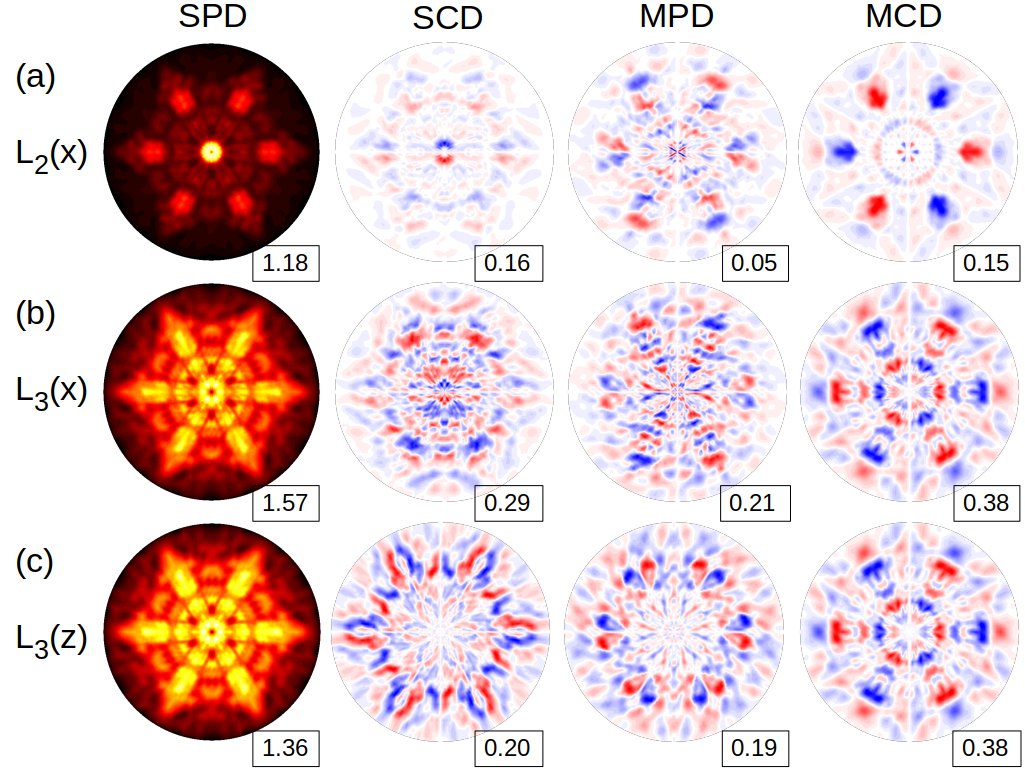}
\caption{Calculated RPED patterns of MnTe(0001) with circular polarized light
  in parallel geometry. 
  (a) L$_2$-resonance ($\hbar\omega$$=$$651.8$eV). ${\bf L}$, ${\bf q}\sim x$.
  (b) L$_3$-resonance ($\hbar\omega$$=$$639.7$eV). ${\bf L}$, ${\bf q}\sim x$.
  (c) L$_3$-resonance. ${\bf L}$, ${\bf q}\sim z$.
  The columns show the four fundamental patterns as defined in section~(i).
  The color scale is always from minimum to maximum intensity,
  which is indicated in the small boxes. In SPD, the minimum
  is black and the maximum is yellow.
  In the other patterns, red is positive, blue is negative and white is zero.
} \label{figS1}
\end{figure}

In the SPD pattern, ${\bf L}$ and ${\bf q}$ are averaged over opposite signs,
so SPD corresponds to a demagnetized sample with random AM domains
and non-polarized light, and it contains no magnetic information.
Accordingly, the patterns have the 6-fold symmetry of the (stepped)
MnTe(0001) surface.
The SCD pattern reflects the common CD in angular distribution in photoemission,
which for X-ray PED is also known as the Daimon effect.
The MPD pattern corresponds to the RPED contrast between two AM domains
with opposite ${\bf L}$ when using unpolarized light.
The fact that this signal is non-zero in Fig.~S1, directly shows that RPED is
sensitive to the sublattice magnetization in AMs.
The MCD pattern is the purely magnetic CD and has been discussed in some
detail in the main text.

In Fig.~S1, the fundamental patterns are shown
for the L$_2$ and the L$_3$-resonance with ${\bf L}$ and ${\bf q}$ along
the $x$ or $z$-axis.
Looking at L$_2(x)$ and L$_3(x)$ in Figs~S1~(a,b)
we see that the SPD pattern has the 6-fold symmetry of the macroscopic
surface.
Since SPD is a symmetric combination of
${\bf L} =\uparrow,\downarrow$ and ${\bf q}=\pm$,
the orientation of ${\bf L}$ and ${\bf q}$ along~$x$ does not break
the symmetry.
The pattern MCD has only 3-fold symmetry and it changes sign under 2-fold
rotation around~$z$, which reflects the point symmetry of
the individual Mn sites ($S_6$).
In the SCD pattern, the 3-fold symmetry is broken by the helicity vector
along~$x$.
The Daimon effect is well seen, especially at the Mn-Mn
forward focussing peak at the center of pattern L$_2$(x)-SCD.
In the MPD pattern the 3-fold symmetry is broken by the N\'eel vector
along~$x$.

Comparing the patterns for L$_2(x)$ and L$_3(x)$, the peak intensities
are quite different, which indicates
that the angular distribution of source wave depends strongly on magnetic
moment of the core-hole ($j=1/2$ or $j=1/2$). However the peak positions
are rather similar, as they appear in both cases at forward
focussing directions, see Fig.~1(c) of the main text.

When comparing the patterns L$_3(x)$ and L$_3(z)$ in Figs.~S1(b,c),
we see that the SPD patterns are very similar.
The small difference is due to the fact that ``unpolarized'' light
is actually polarized in the plane perpendicular to~${\bf q}$.
In contrast, the patterns SCD and MPD are very different between
L$_3(x)$ and L$_3(z)$,
i.e. they depend strongly on the orientation of ${\bf L},{\bf q}$
relative to the crystal axes.

Surprisingly, the MCD patterns for L$_3(x)$ and L$_3(z)$ are
{\em exactly}\/ identical.
So the MCD signal depends only on the relative orientation
between the N{\'eel} vector and the light helicity,
rather than on the orientation of these vectors w.r.t. the crystal axes.
This is a remarkable finding because the axes [1${\bar 1}$00] and [0001]
are not symmetry related.
It shows that the MCD pattern is free of structural CD effects and thus
really measures the magnetic properties.

\subsection*{(iii) Why is the magnetic CD proportional to the structural pattern
  (A$-$B)?}
\begin{figure}
 \includegraphics[width=0.65\columnwidth]{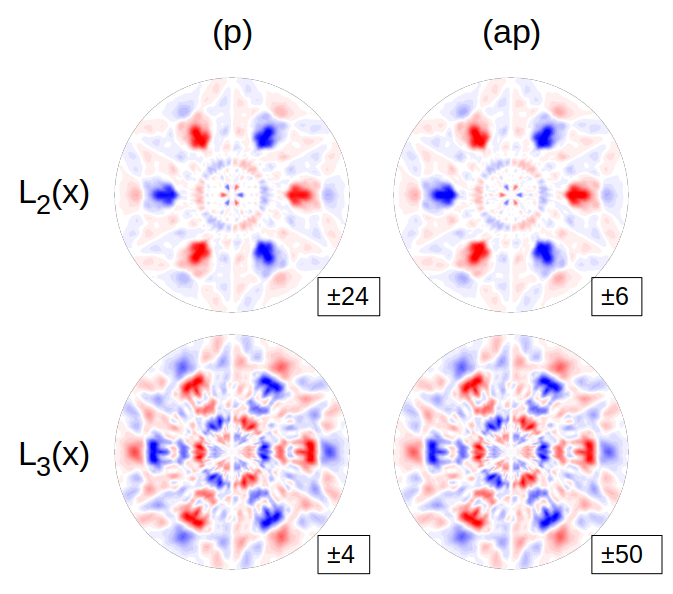}
  \caption{RPED patterns $(p)$ and $(ap)$ as defined in section (ii)
    for MnTe(0001) with circular polarized light at
  the L$_2$ or L$_3$-resonance with
  ${\bf L}$, ${\bf q}\sim x$. For each pattern, the maximum intensity is
  indicated in the small box. The normalization is the same as
  in Fig.4 of the main text.}\label{figS2}
\end{figure}
As seen in Fig.~4 of the main text, the MCD pattern is, up to a constant
intensity scaling, identical with the purely structural pattern (A$-$B).
Here we give an explanation for this important result.
To this end, we decompose the pattern (A$-$B) into two contributions,
with either parallel $(p)$ or antiparallel $(ap)$
alignment between the local spins and the helicity vector. These patterns
are defined as follows.
\begin{eqnarray}
  (p) &=& (a{\uparrow}+)+(a{\downarrow}-)-(b{\uparrow}+)-(b{\downarrow}-)
  \nonumber \\
  (ap) &=& (a{\uparrow}-)+(a{\downarrow}+)-(b{\uparrow}-)-(b{\downarrow}+)
\nonumber \end{eqnarray}
where ($a{\uparrow}+$) is the RPED signal due to emission from sites A only,
with {\em local}\/ spin orientation ${\uparrow}$ and light helicity $+$.
Similarly ($b{\downarrow}-$) corresponds to emission from site B with spin
${\downarrow}$ and helicity~$-$. Note that N\'eel vector ${\bf L}=\uparrow$
corresponds to ($a{\uparrow}$) and ($b{\downarrow}$).
The signals $(p)$ and $(ap)$ are structural RPED patterns
which measure a contrast between emission from sites A and B.
They are invariant under reversal of either N\'eel vector or helicity (or both).
They have the same symmetry 
and are not directly related to the magnetic state
(${\bf L}$) of the system or the light polarization (${\bf q}$).
Therefore we expect them to have the same angular distribution due
to photoelectron diffraction.
The calculated patterns $(p)$ and $(ap)$ are shown in Fig.~S2 for the L$_2$ and
L$_3$ resonances with ${\bf L}$ and ${\bf q}\sim x$.
$(p)$ and $(ap)$ are indeed identical except for the overall intensity
(indicated in the boxes).
We have
\[ \mbox{(A$-$B)} = (p)+(ap) \;, \]
so (A$-$B) has the same symmetry and the same angular distribution as $(p)$ and $(ap)$.
The difference between $(p)$ and $(ap)$ is the relative alignment between
local spin and light helicity, which is parallel in $(p)$ and
antiparallel in $(ap)$.
This changes the local XMCD effect and so
the difference between $(p)$ and $(ap)$ is related to the
XMCD contrast between sublattices A and B.
From the definitions it is easy to see that 
\[ {\rm MCD} = (p)-(ap) \;. \]
Since patterns $(p)$ and $(ap)$ have the same angular dependence,
their sum (A$-$B) and difference (MCD) also have this  same
angular dependence,
which is indeed the case, see Fig.~4 in the main text.

So the reason why the patterns (A$-$B) and MCD are identical, up to 
a constant scaling factor, 
is that they are the sum and difference, respectively,
of two patterns $(p)$ and $(ap)$ with the same angular dependence.
$(p)$ and $(ap)$ have the same angular dependence because
they are purely structural RPED signals with the same symmetry.
They differ only in the relative alignment between ${\bf L}$ and ${\bf q}$,
which determines the overall intensity through the local XMCD effect.

\subsection*{(iv) RPED calculations for bulk MnTe.}
In order to prove that the appearance of the MCD signal is not a surface
effect, we have performed RPED calculations for bulk MnTe using
spherical clusters of 447 atoms with a single Mn emitter site at the
center.
Calculations were done for two clusters, for emission from Mn-A and Mn-B,
and the intensities were summed.
Fig.~S3 shows the results for the L$_3$-resonance
($\hbar\omega$$=$$639.7$eV) with ${\bf L}$, ${\bf q}\sim x$.
The individual patterns
$(\uparrow +)$, $(\uparrow -)$, $(\downarrow +)$, $(\downarrow -)$
are shown, along with the combinations SPD and MCD, defined in section~(i).

\begin{figure}
\includegraphics[width=0.9\columnwidth]{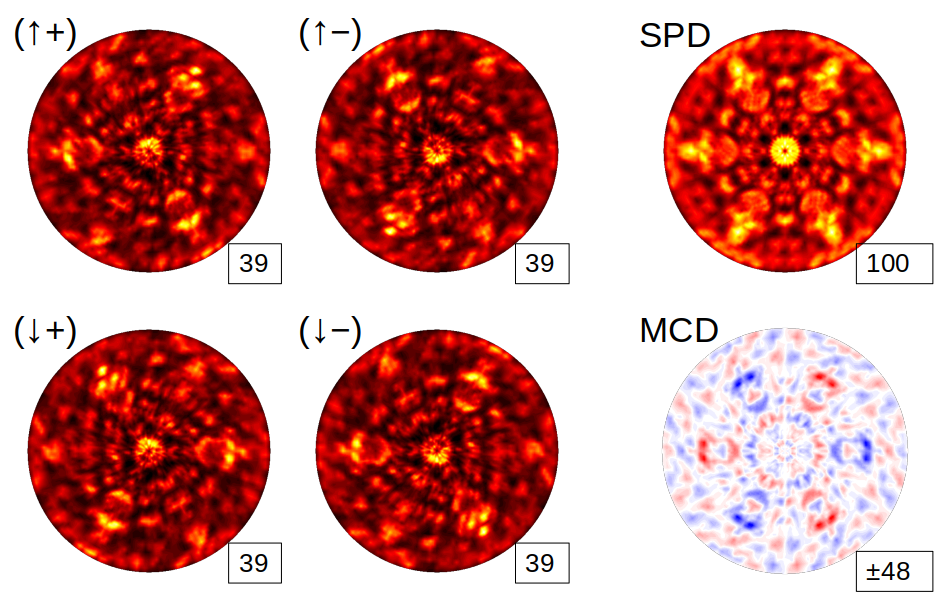}
\caption{RPED patterns for bulk MnTe with circular polarized light at
  the L$_3$-resonance ($\hbar\omega=639.7$~eV) with
  ${\bf L}$, ${\bf q}\sim x$. For each pattern, the maximum intensity is
  indicated in the small box relative to the SPD pattern, whose maximum
  is put to 100.} \label{figS3}
\end{figure}
The patterns are different from those of the MnTe(0001)
surface calculation in Figs.3,4(a),S1(b).
The bulk patterns in Fig.~S3 display
more fine structure and sharper diffraction peaks, which is expected because
photoelectrons emitted from a bulk site experience more scattering events.
Importantly however, the symmetry of all patterns is the same for bulk and
surface, and the positions of the dominant peaks is similar.
The amplitude of the MCD signal
is  comparable with that of the surface,
considering the more homogeneous intensity distribution of the latter.
These results clearly show that the magnetic CD in RPED of altermagnets
is not a surface effect but essentially a bulk property.

\subsection*{(v) Comparison with a conventional antiferromagnet (MnO)}
\begin{figure}
\includegraphics[width=\columnwidth]{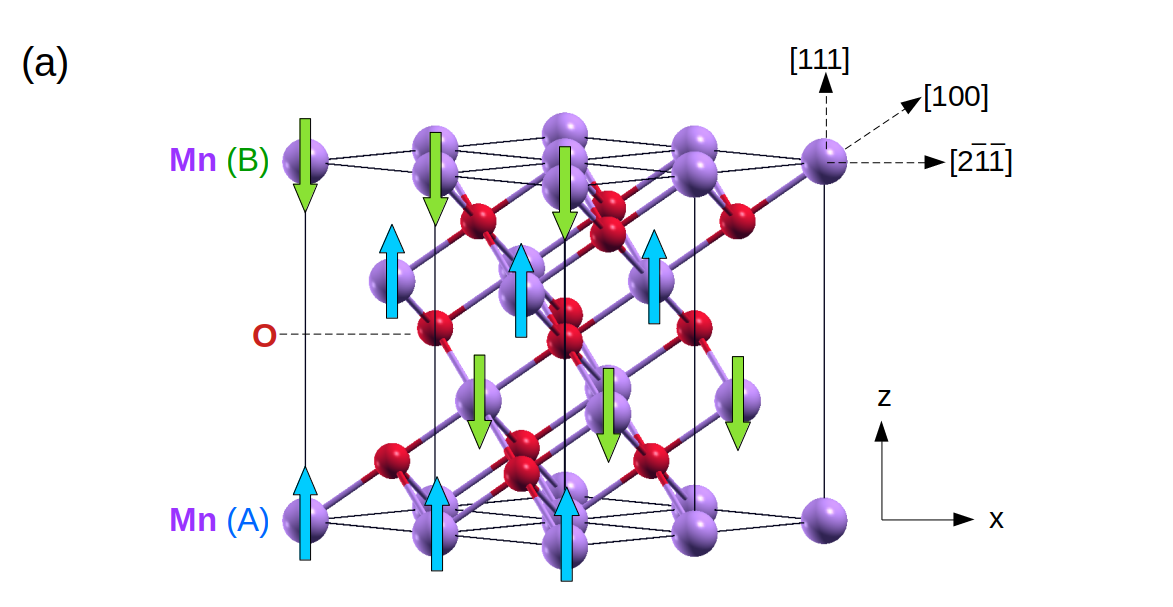}
\includegraphics[width=\columnwidth]{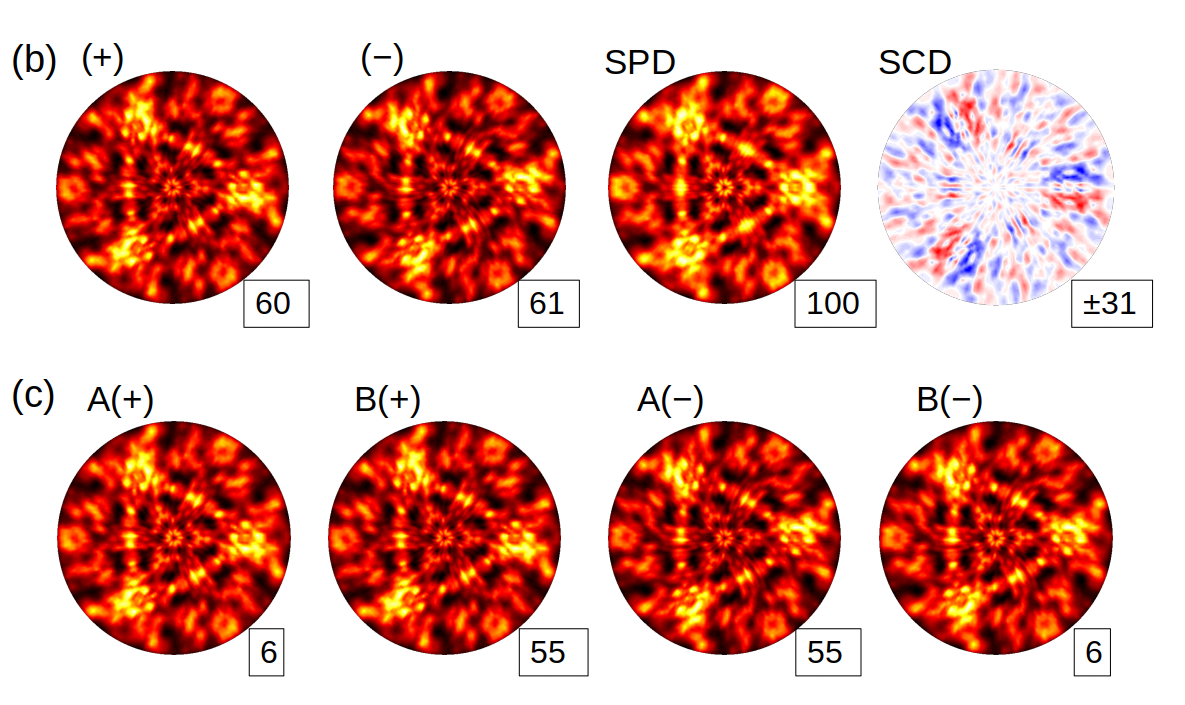}
  \caption{
(a) Ball-and-stick model of bulk MnO with the [111] axis oriented
    along~$z$. The antiferromagnetic ground state with ferromagnetically
    aligned (111) layers is shown.
    The N\'eel vector is taken ${\bf L}\sim z$.
    Mn(A) and Mn(B) denote the spin-up and spin-down sublattice,
    respectively.
(b)
  RPED patterns for bulk MnO with circular polarized light at
  the maximum of the Mn L$_3$-resonance ($\hbar\omega=639.7$~eV)
  with ${\bf L}$, ${\bf q}\sim [111] = z$.
  (+) and ($-$) correspond to right and left circular
  polarized light. SPD is the sum and SCD is the difference
  of (+) and ($-$).
  For each pattern, the maximum intensity is
  indicated in the small box relative to the SPD pattern, whose maximum
  is put to 100.
(c) Decomposition of the patterns (+) and ($-$) into their
  contributions from sites A and B.
} \label{figS4}
\end{figure}

Here we present RPED calculations for the conventional
antiferromagnet MnO. MnO has rocksalt structure and an antiferromagnetic
order with ferromagnetically aligned (111) planes whose
spin direction alternates along [111].
The sublattice magnetization is in-plane~\cite{roth58},
but for simplicity we take the N\'eel vector ${\bf L}$
out-of-plane along the [111] axis, see Fig.~\ref{figS4}a.
As in section (iv) the calculation is done for an oriented crystal in the bulk.
We used a spherical cluster of 389 atoms with a single Mn emitter at
the center.
Fig.~\ref{figS4}~b shows the calculated RPED patterns
at the maximum of the Mn L$_3$-resonance ($\hbar\omega=639.7$~eV)
in parallel geometry, i.e. with ${\bf L}$, ${\bf q}\sim z$.
(+) and ($-$) denote right and left circular polarization.
SPD is the sum of (+) and ($-$) and SCD is the difference.
Fig.~\ref{figS4}~b gives the contributions from the two sublattices A and B.
(Note that these single-sublattice contributions are not measurable.)

All the patterns have three main peaks which correspond to
the Mn$\rightarrow$O nearest-neighbor
forward focussing directions. 
This is similar to the main PED peaks from Mn-A sites in MnTe, which
are caused by Mn-A$\rightarrow$Te focussing (see Fig.~1c of the main text).
Indeed, with the chosen crystal orientations, the Mn-A sites in MnTe
and {\em all}\/ Mn sites in MnO have an octahedral environment with
the same orientation.
The crucial difference between the altermagnet MnTe and
the conventional antiferromagnet MnO is that in MnO, all octahedra
have the same orientation, while in MnTe, the octahedra of the Mn-B sites
are rotated by $\pi$ around~$z$, giving a totally different PED pattern
(Fig.~1c and Fig.~4b,c in the main text).

The RPED patterns of MnO in Fig.~\ref{figS4}~c all look very
similar, but the total intensity (indicated in the small boxes)
depends strongly on relative alignment
between local magnetic moment ${\bf m}$ and the light helicity ${\bf q}$.
For A($+$) and B($-$) the alignment is parallel (${\bf m}\cdot {\bf q}>0$)
and the intensity is weak. For A($-$) and B($+$) with antiparallel
alignment, the intensity is almost 10 times larger.
As already explained for MnTe in the main text,
the reason for this strong intensity difference is the
large negative XMCD effect at the Mn L$_3$-edge, where
X-ray absorption (and thus also resonant photoemission) is maximum
when ${\bf m}$ and ${\bf q}$ are antiparallel.

The structural CD (pattern SCD) is large with $\pm 31$\%.
The MCD signal (not shown) is exactly zero.
This can be understood as follows.
Upon time-reversal, i.e. when the N\'eel vector ${\bf L}$
and thus all the Mn spins are reversed,
the RPED patterns of sublattices A and B are interchanged,
e.g. pattern A(+) becomes B(+).
This does not change any experimentally measurable pattern,
where the signals from the two sublattices are summed.
Therefore, time-reversal does not change any (experimentally measurable)
RPED pattern. As a consequence, the magnetic CD, defined as
MCD$=$CD(${\bf L}$)$-$CD($-{\bf L}$)
is exactly zero. For the same reason, the MPD signal is also exactly zero.

In a compensated magnetic structure, time-reversal
is equivalent to interchanging the two magnetic sublattices A and B.
In a conventional antiferromagnet, this transformation is a translation
by a Bravais lattice vector and does not change the crystal orientation.
As a consequence, the RPED patterns of two sublattices A and B                 
are simply swapped (rather than rotated as in an altermagnet).
Then, any experimental RPED signal (which is the sum of A and B)
is time-reversal even.
It follows that the magnetic signals MCD and MPD, which are
time-reversal odd by definition, must vanish exactly.

\bibliography{mnte}{}